\documentclass{optica-article}

\journal{opticajournal} 

\articletype{Research Article}

\usepackage{lineno}

\usepackage{caption}
\usepackage{subcaption}
\usepackage[per-mode = symbol]{siunitx}
\usepackage{tikz}

\usepackage{xcolor}

\begin{document}

\title{Optimized thermal control of a dual-wavelength-resonant nonlinear cavity}

\author{Fabian Meylahn \authormark{1,2,*}, Henning Vahlbruch\authormark{1,2}, and Benno Willke\authormark{1,2}}

\address{\authormark{1}Max Planck Institute for Gravitational Physics (Albert Einstein Institute), D-30167 Hannover, Germany\\
\authormark{2}Leibniz University Hannover, D-30167 Hannover, Germany}
\noindent

\email{\authormark{*}fabian.meylahn@aei.mpg.de}

\begin{abstract*}

Optical resonator-enhanced nonlinear interactions are of great importance for the efficient generation of continuous-wave second harmonic generation, optical parametric oscillation, frequency mixing, and the generation of squeezed light. In order to maximize these interactions within the intra-cavity nonlinear material, high intensities, optimal phase matching, and simultaneous resonance of all interacting fields are required. However, the dispersion of the optical resonator often prevents the co-resonance of multiple wavelengths. Here, we present a novel implementation using a monolithic bimetallic heat sink for controlling the resonator dispersion based on a shallow temperature gradient directly applied to a section of the nonlinear crystal. 
This method enables precise dispersion control and is designed to minimize mechanical and thermal stresses in the nonlinear crystal, thus providing an additional method for designing highly efficient and reliable resonator-enhanced nonlinear devices for demanding applications such as gravitational wave detection, quantum optics, and frequency conversion.

\end{abstract*}

\section{Introduction}

Nonlinear optics enables wavelength conversion of laser light and the generation of quantum correlations, thus providing important tools for overcoming the limits of quantum noise in precision measurements \cite{Caves1981,GEO2011a,Schnabel2017} and generating non-classical states for quantum communication \cite{Leuchs2014, Gehring2015}.
Since 2010, squeezed states of light generated through nonlinear interactions have enhanced gravitational wave detection sensitivity beyond classical limits, namely shot noise and radiation pressure noise \cite{GEO2011a,Grote2013,Tse2019, VirgoSQZ, VirgoQuantumBackaction, LigoSQZ,Jia2024}.
These quantum states are generated by cascaded resonator-enhanced processes: Light at the fundamental wavelength of \SI{1064}{\nm} for current gravitational wave detectors undergoes frequency doubling to generate a pump field of \SI{532}{\nm}. This pump then drives a parametric down-conversion in a second resonator containing a nonlinear medium, and generates thereby quantum correlations.
The development of nonlinear resonators generating squeezed vacuum states of light for the application in gravitational wave observatories has led to several architectural variants: A linear semi-monolithic cavity design achieved a non-classical noise reduction of \SI{15}{\dB} without active dispersion control \cite{Vahlbruch2016}, representing the current performance benchmark. Long-term operation was demonstrated at the GEO600 gravitational wave detector with a single-resonant semi-monolithic resonator design \cite{GEOSqueezer,Khalaidovski2012}. 
Since 2018 the Virgo gravitational wave detector sensitivity is squeezed light enhanced using a doubly-resonant semi-monolithic resonator design with segmented crystal heating for dispersion control \cite{VirgoSqueezer}, while the LIGO detectors, on the other hand, employ doubly resonant bow-tie traveling wave resonators that control dispersion by mechanically adjusting the positions of a wedged nonlinear crystals 
\cite{Chua:11, ANUbowtie2016, LIGOSQZ2016}. 
At the wavelength of \SI{1550}{\nm}, multi-temperature-zone heating was introduced and demonstrated as a dispersion-control technique for nonlinear resonators generating squeezed states of light, including doubly resonant semi-monolithic and monolithic resonators \cite{Schoenbeck2018a,Schnabel2019,Hagemann2024}.

Compared to single resonant resonator designs, double resonant cavity designs generally offer the advantages of reduced external pump power requirements and the additional possibility of stabilizing the resonator length by detecting the second harmonic wavelength. This allows, for example, squeezed vacuum states of light to be generated without coherent excitation at the fundamental wavelength. However, a prerequisite for effectively implementing this process is the simultaneous resonance of all wavelengths inside the resonator.

Building on our previous work with astigmatism-corrected bow-tie resonators at \SI{1550}{\nm} \cite{Meylahn2022a}, which demonstrated compatibility with next-generation detectors \cite{ET_Maggiore_2020}, we present here a doubly resonant bow-tie resonator for the generation of squeezed light at \SI{1064}{\nm}.

Our design uses a novel monolithic bimetallic heat sink with a controlled longitudinal temperature gradient for dispersion control that has not yet been described in the literature.
The heat sink, milled in a single piece, provides a continuous mechanical contact surface for the nonlinear crystal, creating a uniform and shallow temperature gradient within the crystal and thus minimizing thermal stresses.
This monolithic design further reduces mechanical stress and relaxes manufacturing tolerances, as it eliminates the need for a critical alignment of separate heating segments. Additionally, the controlled thermal gradient ensures a more homogeneous temperature distribution perpendicular to the direction of light propagation, effectively preventing beam distortion and suppressing transverse fluctuations in the parametric gain.
Our design is also compatible with the realization of semi-monolithic or monolithic cavities, unlike wedge-shaped crystal tuning methods \cite{Stefszky2011,ANUbowtie2016}, because shifting a wedged crystal is not possible, when at least one side of the crystal need to act as a cavity mirror.

In this paper, we first describe the thermal design of our bimetallic heat sink and validate its performance through thermal radiation imaging. We then perform a complete optical characterization of the bow-tie resonator with an embedded gradient-controlled nonlinear crystal and map the double resonance and parametric amplification over the entire operating temperature range. Finally, we discuss the relationship between the measured parametric amplification and the achievable quantum noise reduction (squeezing strength).

\section{Setup and simulation}
\label{sec:setupAndSim}

\begin{figure}[htbp]
\centering\includegraphics[width=0.6\linewidth]{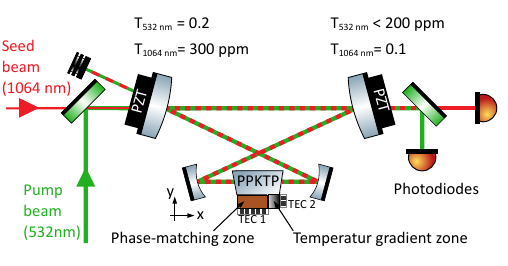}
\caption{Configuration of the bow-tie cavity, for dispersion-control enabled co-resonance and, thus, for high nonlinear interaction. The periodically poled potassium titanyl phosphate (PPKTP) crystal is mounted on a bimetallic heat sink, creating a constant-temperature zone and an adjustable thermal gradient zone via thermoelectric coolers (TECs). Two piezoelectric elements (PZTs) attached to the convex coupling mirrors enable resonator length tuning. A bright \SI{532}{\nm} pump beam and a weak \SI{1064}{\nm} seed beam (at half the pump-light frequency) are combined on a dichroic beam splitter and injected through the same input port. This port exhibits a power transmission~(T) of 300 parts per million (ppm) at the seed wavelength.}
\label{fig:setup}
\end{figure}

The temperature control scheme for a nonlinear optical medium presented here is demonstrated in a dual-wavelength resonant traveling wave  cavity, which is optimized for the generation of squeezed light by featuring an astigmatism-compensated design according to \cite{Meylahn2022a}. The rigid spacer bow-tie cavity, as shown in Fig.~\ref{fig:setup}, consists of two convex mirrors with a radius of curvature of \SI{1}{\m}. These mirrors, which can be displaced via piezoelectric elements (PZT), serve as input and output coupling mirrors for the \SI{532}{\nm} pump beam and the \SI{1064}{\nm} fundamental wavelength beam, respectively. Two concave mirrors focus the beams into a periodically poled nonlinear potassium titanyl phosphate (PPKTP) crystal. In order to achieve a waist radius of \SI{38}{\um} for the fundamental field and \SI{27}{\um} for the second harmonic field, the radius of curvature of the highly reflective mirrors was chosen to be \SI{-0.1}{\m}.

The mirrors and nonlinear crystal form a resonator with an optical round-trip length of \SI{64}{\cm}, corresponding to a free spectral range of \SI{466}{\MHz}. For the pump field at \SI{532}{\nm}, the input coupler has a power transmission of \SI{20}{\percent}. Combined with the high reflectivity of the other resonator mirrors at this wavelength, this yields a pole frequency of \SI{8.5}{\MHz}. The output coupler for the fundamental field at \SI{1064}{\nm} has \SI{10}{\percent} power transmission, which in combination with the much lower transmission of all other cavity mirrors results in a pole frequency of \SI{4.1}{\MHz}.

The laser fields injected into the resonator are obtained from a non-planar ring oscillator laser source at a wavelength of \SI{1064}{\nm} and a subsequent in-house build second harmonic generation unit. This ensures that the cavity input field at the second harmonic has exactly twice the frequency of the fundamental field, which is used to probe the parametric amplification factor introduced by the nonlinear crystal inside the resonator. 
The bow-tie design ensures largely collimated beams for both the input and output fields, enabling long propagation paths without large-diameter optics. At the fundamental wavelength, the output beam has a waist diameter of \SI{239}{\um}, located between the convex mirrors.

The PPKTP crystal, with a length of \SI{11.5}{\mm} and anti-reflective coated wedge-shaped interfaces, is mounted on a bimetallic heat sink. A section measuring \SI{8.5}{\mm} encompasses the centrally located beam waist and rests on the copper part of the heat sink, which is stabilized at the crystal phase-matching temperature using a thermoelectric cooler (TEC~1) in conjunction with a in-house build high-precision temperature controller. The remaining \SI{3}{\mm} part of the crystal sits on the stainless steel part of the heat sink, which is soldered and screwed to the copper block before the final milling step of the entire heat sink assembly. Along this stainless steel part and consequently along the second part of the crystal, a second thermoelectric cooler (TEC~2) creates a temperature gradient. This gradient spans between the  cooper part and the right side of the stainless steel part of the heat sink and is controlled by an additional high-precision controller.

The difference in thermal conductivity between the heat sink materials determines the temperature distribution. The copper part (\SI{401}{\W\per\m\per\K}, \cite{NIST_Cu_thermal_cond}) has negligible thermal gradients due to its high conductivity compared to the 1.4301 stainless steel part \SI{15}{\W\per\m\per\K}, \cite{Peet_2011_steel_thermal_cond}), which, due to its lower conductivity, develops a significant thermal gradient in the beam propagation direction (x-direction in Fig.~\ref{fig:setup}) when heated or cooled from the right side.
The PPKTP crystal with a thermal conductivity of \SI{13}{\W\per\m\per\K} \cite{Steinlechner2013_PPKTP} is in contact to the heat sink via a thin, soft, highly thermally conductive layer of Indium foil. The crystal adapts both to the constant temperature zone and to the temperature gradient generated by the heat sink. The the other three long surfaces of the crystal are encapsulated by insulation material with low thermal conductivity ($<$\SI{0.3}{\W\per\m\per\K}), which minimizes heat flow through those surfaces and thus minimizes the temperature gradient in the crystal perpendicular to the direction of beam propagation.

\begin{figure}
     \centering
     
     \begin{subfigure}[b]{0.8\textwidth}
        \centering\includegraphics[width=\linewidth]{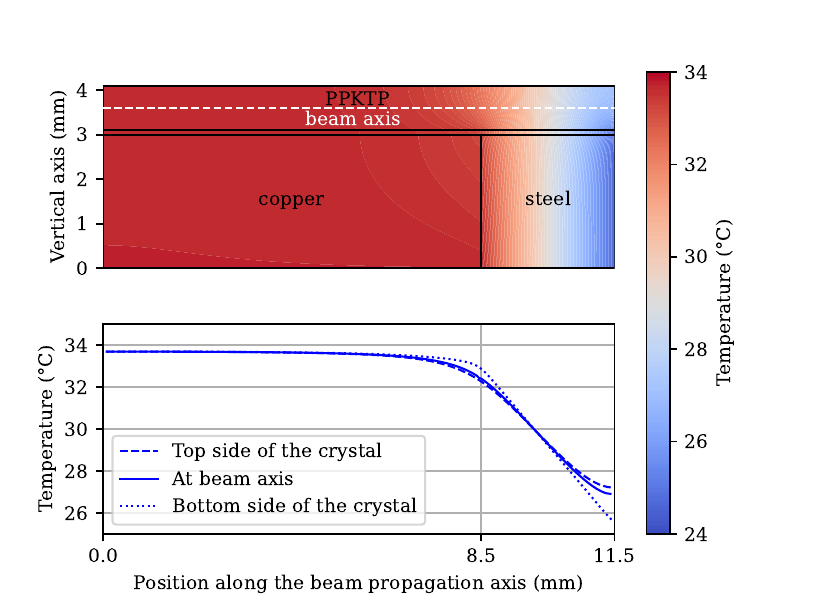}
        \caption{ }
         
     \end{subfigure}
     \hfill
     \begin{subfigure}[b]{0.8\textwidth}
        \centering\includegraphics[width=\linewidth]{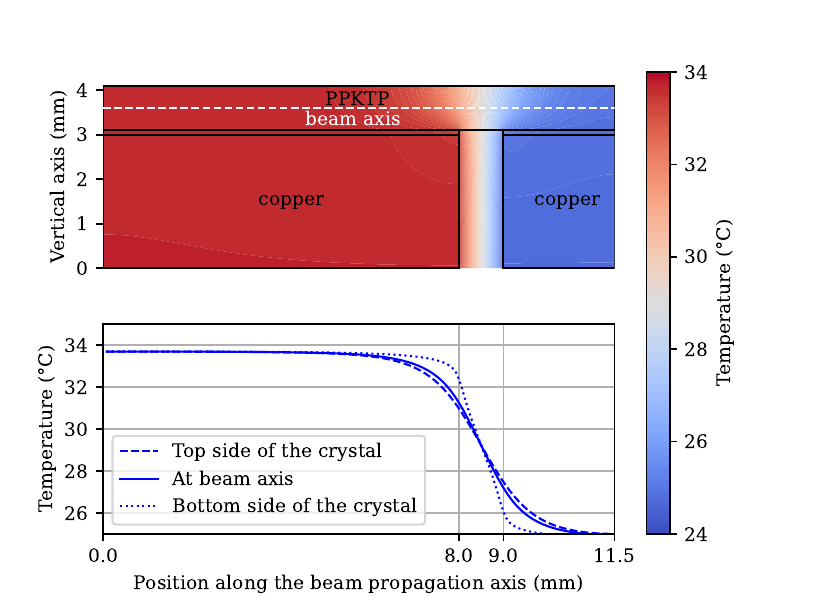}
        \caption{ }
        
     \end{subfigure}
     
        \caption{Finite element simulations of temperature distributions in the heat sink and the nonlinear PPKTP crystal.\\ \textbf{a)} Gradient heating: A controlled temperature gradient applied to the right heat sink surface propagates through the bimetallic construction to the crystal. Primary heat flow occurs through the heat sink and the generated temperature gradient is adapted by the crystal\\\textbf{b)} Segmented heating: The crystal itself forms the only thermal bridge across an air gap between two temperature zones. Heat flow through the crystal creates a strong thermal gradient that peaks at the air-gap position and close to the heat sink interface. }
        \label{fig:sim}
\end{figure}

The thermal configuration described here was simulated using Elmer FEM  software package~\cite{ELMER,malinen2013elmer} and is shown in Fig.~\ref{fig:sim}a. For comparison, Fig.~\ref{fig:sim}b shows an implementation of the previously used segmented heating scheme \cite{Schoenbeck2018a,VirgoSqueezer,Meylahn2022b}. In the bimetallic heat sink design (Fig.~\ref{fig:sim}a), the fragile nonlinear crystal is fully supported by the heat sink along its entire length. Conversely, in segmented heating with two air gap separated temperature zones (Fig.~\ref{fig:sim}b), the crystal lacks support over the air gap of \SIrange{0.5}{1}{\mm}. This requires tight manufacturing tolerances for both heat sink segments in order to avoid mechanical stress on the crystal. Otherwise, height differences could lead to stresses sufficient to break the crystal or at least cause stress-induced changes in the refractive index.
In gradient heating, heat flows through the steel part of the heat sink, causing a controlled temperature gradient, but also requiring higher TEC heating/cooling power. This gradient in the heat sink is transferred to the nonlinear crystal via the thin, highly conductive Indium layer. In segmented heating, the crystal itself forms the only thermal bridge between the zones and creates a steep temperature gradient that reaches its maximum value at the bottom of the crystal (dotted graph in Fig.~\ref{fig:sim}b).

\section{Results and Discussion}

We first describe a thermal characterization of the heat sink in subsection~\ref{subsec:thermal} to verify the simulation results discussed in section \ref{sec:setupAndSim}. Following this validation, subsection~\ref{subsec:optic} presents the optical characterization of the complete resonator system with the embedded nonlinear crystal.

\subsection{Thermal characterization}
\label{subsec:thermal}

\begin{figure}[htbp]
\centering\includegraphics[width=0.9\linewidth]{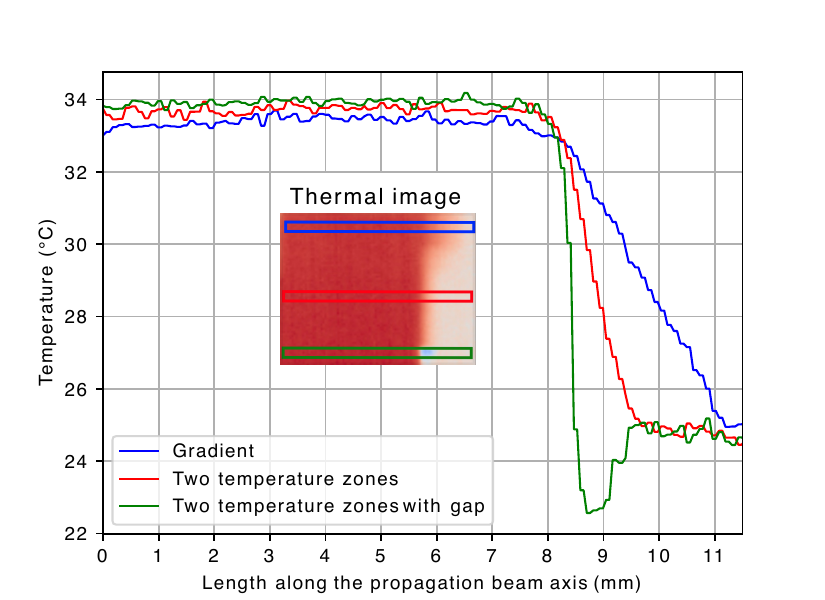}
\caption{Measured temperature distribution along the beam propagation axis for different heating schemes (inset images). The phase-matching constant-temperature zone is maintained at \SI{33.7}{\celsius} in all configurations. The gradient heating scheme exhibits a \SI{2.7}{\celsius\per\mm} thermal gradient. For segmented heating, a steep thermal slope develops across the thermal isolation layer here simulated as an air gap when bridged by the thin polyamide layer, with an even steeper gradient observed without this interfacial layer. The second temperature zone in segmented heating is set to \SI{24.7}{\celsius}. Thermal images were corrected for a slight spurious gradient caused by camera temperature non-uniformity, verified through camera rotation tests.}
\label{fig:thermal_image}
\end{figure}

To validate the simulations in Fig.~\ref{fig:sim} experimentally, we took thermal images of the heat schemes (Fig.~\ref{fig:thermal_image}). The image of the heat sink show areas that exhibit characteristic thermal properties for both the gradient and segmented heating schemes. For accurate temperature measurement, a thin layer of polyamide was applied to normalize the surface emissivity of the metal materials. The temperatures of the heat sink segments were stabilized and monitored using negative temperature coefficient thermistors embedded in the heat sink.
The measured thermal profile along the beam propagation axis, which was derived from image processing, largely corresponds to the simulated distributions in Fig.~\ref{fig:sim}.

\subsection{Optical characterization}
\label{subsec:optic}

\begin{figure}[htbp]
\centering\includegraphics[width=0.8\linewidth]{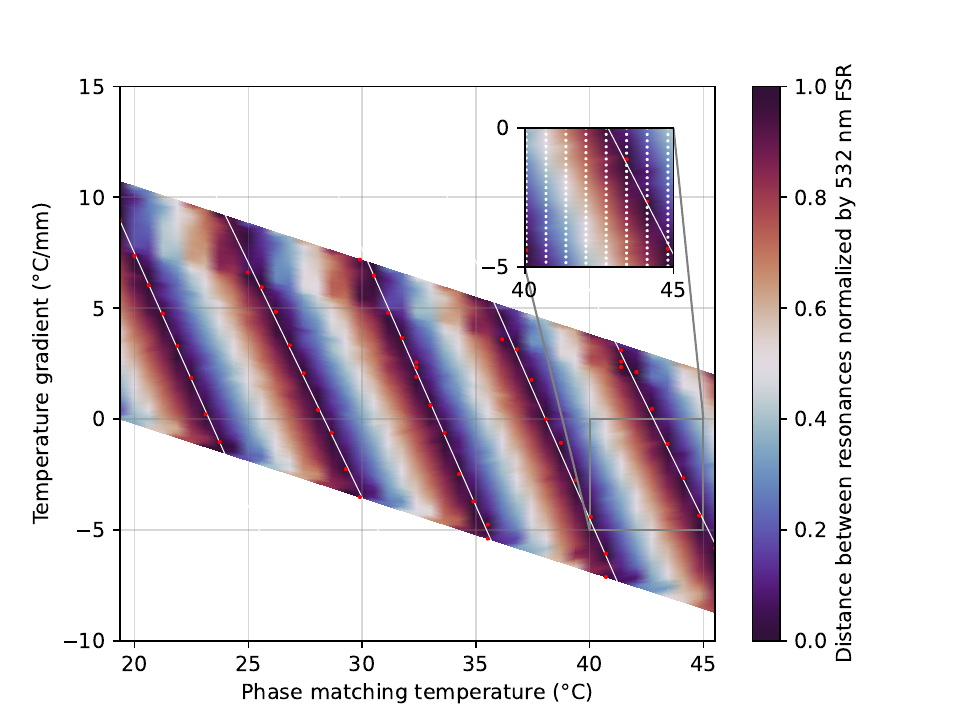}
\caption{Plot showing the distance between the fundamental resonances of the TEM$_{00}$ mode of \SI{1064}{\nm} light from \SI{532}{\nm} light, normalized by the free spectral range (FSR) of the \SI{532}{\nm} light, color-coded as a function of the phase matching temperature and temperature gradient settings. White lines represent linear fits to  co-resonance conditions (red dots). The results show a required temperature gradient along the steel section of \SIrange{-2.07}{-2.32}{\celsius\per\mm} per degree change in phase matching temperature (temperature of the cooper section). The inset shows measurement points (white) with red markers indicating the positions closest to the co-resonance.}
\label{fig:double_reso}
\end{figure}

To prevent water condensation, one boundary condition is that no heat sink component may fall below \SI{19}{\celsius}. At the other end, due to the finite heating power, the constant temperature zone could not be heated above \SI{46}{\celsius} and the stainless steel end surface could not be heated above \SI{52}{\celsius}. The resulting parameter space, which is spanned by the phase matching temperature applied to the copper part of the bimetallic heating element and the temperature gradient in the stainless steel part, was mapped during the scanning of the resonator length and the detection of the resonance positions for both wavelengths.
We determined the distance between the fundamental resonances of the TEM$_{00}$ mode of \SI{1064}{\nm} light and \SI{532}{\nm} light, normalized by the free spectral range of the \SI{532}{\nm} light, and represented it as color coding in the two-dimensional parameter range in Fig.~\ref{fig:double_reso}.

During post-processing, we identify and mark the data points closest to the double resonance with red dots in Fig.~\ref{fig:double_reso}. The fit of linear functions to the dot positions, shown as white lines, shows that the temperature gradient must decrease by \SI{-2}{\celsius\per\mm} per \SI{1}{\celsius} increase in phase matching temperature to keep the co-resonance condition.
The scan confirms that under these boundary conditions, a suitable temperature gradient does always exists to achieve co-resonance.

To quantify the interaction strength inside the nonlinear crystal, we measure the optical parametric gain by feeding a laser field with the fundamental wavelength and with \SI{21}{\uW} power into the resonator through the \SI{532}{\nm} coupling mirror (Fig.~\ref{fig:setup}). The mode overlap with the fundamental resonator eigenmode is \SI{65}{\percent}, which means that only this fraction of the injected light field interacts with the second harmonic field. The second harmonic beam is injected through the same coupling mirror with an input power of \SI{14.3}{\mW} and a mode overlap of \SI{83}{\percent}. This results in an effective power of \SI{14}{\uW} for  the fundamental field coupled to the fundamental mode of the resonator and, correspondingly, \SI{11.8}{\mW} for the pump field.

\begin{figure}[htbp]
    \centering
    \includegraphics[width=0.8\linewidth]{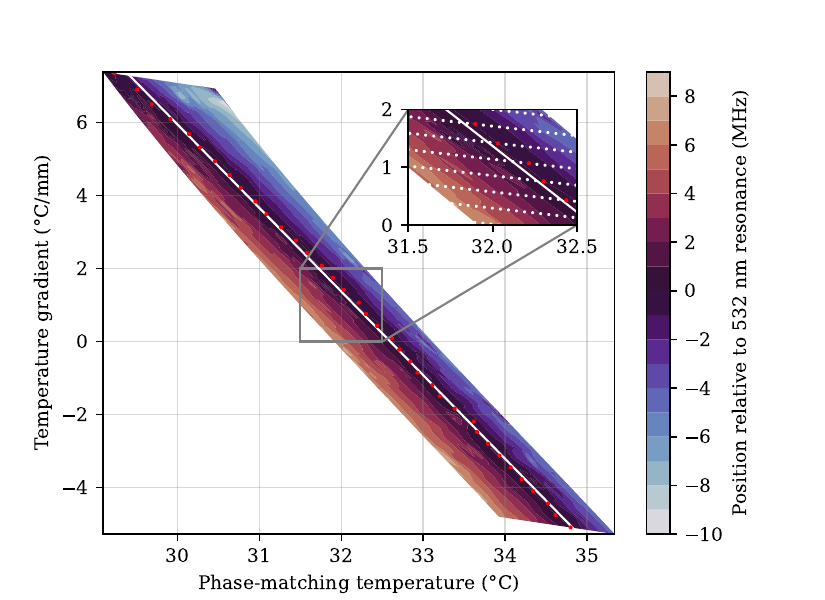}
    \caption{
    Detailed mapping of double-resonance conditions across phase-matching temperatures from \SIrange{29}{35}{\celsius}. The inset displays measurement data points (white) with red markers indicating positions closest to co-resonance. Co-resonance conditions exhibit a near-linear dependence on temperature gradient, with the white fit line showing a slope of \SI{-2.29}{\celsius\per\mm\per\celsius}.
    }
    \label{fig:double_reso_zoom}
\end{figure}

\begin{figure}[htbp]
         \centering\includegraphics[width=0.7\linewidth]{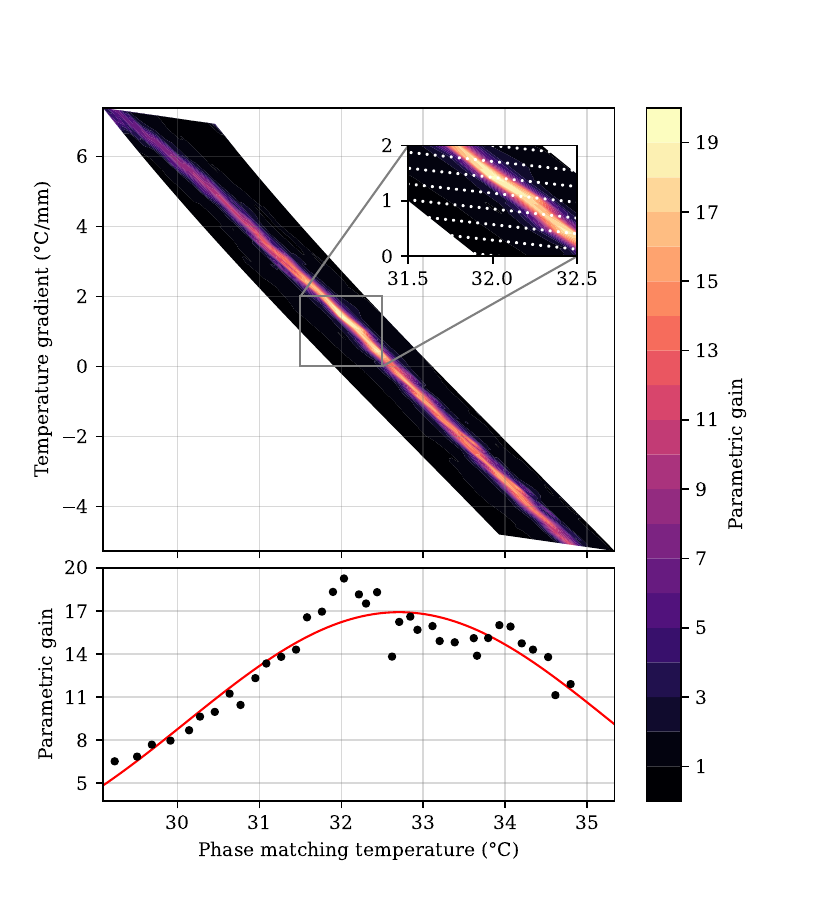}
        

    \caption{\textbf{Upper: }Parametric gain color coded versus phase-matching temperature and temperature gradient. The inset displays the measurement step size (white points).\\
    \textbf{Lower: }Parametric gain values per phase-matching temperature. The fitted model yields a  peak gain of 16.91 at an optimal temperature of \SI{32.7}{\celsius} with a full width at half maximum of \SI{6.3}{\celsius}}
    \label{fig:gain}
\end{figure}

By manually optimizing maximum optical parametric gain, we identify the phase matching temperature range shown in Figures~\ref{fig:double_reso_zoom} and \ref{fig:gain}. Figure~\ref{fig:double_reso_zoom} shows a contour plot representing detailed double resonance conditions, with optimal resonance points marked as red dots. A linear fit to these points agrees with the results in Fig.~\ref{fig:double_reso} and confirms the required gradient fit of \SI{-2.29}{\celsius\per\mm} per \SI{1}{\celsius} change in phase matching temperature. The color scale represents the frequency deviation of the second harmonic resonance from the resonance of the fundamental field, with the color code zoomed in on the range where the deviation does not exceed the \SI{8.5}{\MHz} pole frequency, beyond which efficient nonlinear interaction becomes impossible.

During the same measurement sequence, we modulated the phase of the second-harmonic field to continuously sweep between parametric amplification and deamplification. For each data point, we evaluated 30 measurements to determine the maximum seed field amplification, normalized by the seed transmission without pump field. This procedure yields the parametric gain shown in Fig.~\ref{fig:gain}.

A comparison of the co-resonance range in Fig.~\ref{fig:double_reso_zoom} and the range of maximum parametric gain in the upper plot in Fig.~\ref{fig:gain} shows considerable overlap. As expected, strong nonlinear interactions only occur under co-resonance conditions, which is consistent with their intensity-dependent scaling \cite{Gerry2004}.
The highest parametric gain occurs near \SI{32}{\celsius}, as can be seen in Fig.~\ref{fig:gain}. The lower plot in Fig.~\ref{fig:gain} shows the peak gain values for each phase matching temperatures. An expected profile shape is fitted to the measurement points (red curve) \cite{photonics11010040}. We observe gains of up to 19 at a mode-corrected pump power of \SI{11.8}{\mW}.

The measured optical parametric gain $G$ refers to the pump parameter $x = \sqrt{P/P_{\text{thr}}},$ where $P$ is the pump power and $P_{\text{thr}}$ is the parametric oscillation threshold, according to \cite{Chelkowski2007_PhD}:
 
\begin{align}  
G &= \frac{1}{(1 - x)^2}  
\label{eq:1}  
\end{align}

The pump parameter $x$ is typically used to describe the operation of an optical parametric amplifier operating below its threshold power $P_{\text{thr}}$. This parameter can be used to calculate the achievable quantum noise reduction, also known as squeezing, through optical parametric amplification.

For a classical gain of $G = 19$, equation~\ref{eq:1} implies a pump parameter of $x = 0.77$. The non-classical noise reduction is then modeled using variances of the squeezed state:
\begin{align}
\Delta^2\hat{X}_{\pm} &= 1 \pm \eta_{\text{tot}} \frac{4x}{(1 \mp x)^2 + (f/f_p)^2}
\label{eq:2}
\end{align}
Here, $ \Delta^2\hat{X}_{-} $ and $\Delta^2\hat{X}_{+}$ denote the squeezed and anti-squeezed quadrature variances, respectively, $f_p$ is the pole frequency of the resonator at the squeezed wavelength, and $\eta_{\text{tot}}$ represents the total quantum efficiency from generation to measurement \cite{Vahlbruch2016}.

Based on the pump parameter of $x=0.77$ and an assumed total quantum efficiency of $\eta_{tot}=\SI{97.5}{\percent}$, these relationships predict an anti-squeezing rate of \SI{17.4} dB and a squeezing rate of \SI{13.8}{\dB} for measurements well below the resonator pole frequency and without significant phase noise contributions during the squeezed field detection. The feasibility of these parameters is supported by three key factors: the proven low-loss operation of similar bow-tie resonators at other wavelengths \cite{Meylahn2022a}, the availability of photodiodes with very high quantum efficiency at \SI{1064}{\nm} \cite{Vahlbruch2016}, and the inherent advantages of bow-tie resonator designs, including backscatter immunity \cite{Chua:11} and collimated beam output.

\section{Conclusion}

To the best of our knowledge, this is the first report in literature describing a thermal-gradient approach based on a monolithic bimetallic heat sink for dispersion control in nonlinear resonators. This technique allows for simultaneous phase matching and optical co-resonance without auxiliary components, minimizing optical losses and enabling continuous mechanical support along the full crystal length within a compact unit. 
Our design is fully compatible with linear, bow-tie, and monolithic resonator architectures, offering enhanced mechanical robustness and reduced mechanical and thermal stress on the crystal.
The heating scheme was successfully demonstrated in a double-resonant nonlinear bow-tie resonator at a wavelength of \SI{1064}{\nm}, achieving high optical parametric gain, which highlights its potential for generating highly squeezed states of light. 
Notable practical benefits of our approach include relaxed manufacturing and assembly tolerances by milling the heat sink as a single and rigid piece, enhanced stability by employing a single monolithic heat sink, minimal thermo-optical distortion, and effective suppression of parametric gain fluctuations and beam deformation by applying a shallow temperature gradient. Beyond squeezed light generation, the concept extends to sum- and difference-frequency generation in multi-resonant systems. By eliminating moving parts and enabling full mechanical support of the nonlinear crystal, the technique provides a robust, low-loss foundation for advanced applications such as gravitational wave detection, quantum communication, and precision metrology. Additionally, our thermal-gradient approach offers an ideal platform for experimentally exploring enhanced nonlinear interactions under strong focusing conditions, as proposed in \cite{Lastzka07}.

\bibliography{referencesCollaborations}

\end{document}